\documentclass{article}

\PassOptionsToPackage{sort&compress}{natbib}

\usepackage[main, preprint]{neurips_2026}

\setcitestyle{numbers,square}

\usepackage{amsmath}
\usepackage{amsfonts}
\usepackage{amssymb}
\usepackage{graphicx}
\usepackage{hyperref}
\usepackage{booktabs}
\usepackage{subcaption}    %
\usepackage{longtable}
\usepackage{pifont}  %

\usepackage[capitalize]{cleveref}

\crefname{section}{Sec.}{Secs.}
\Crefname{section}{Section}{Sections}

\crefname{table}{Tab.}{Tabs.}      %
\Crefname{table}{Table}{Tables}    %

\crefname{figure}{Fig.}{Figs.}     %
\Crefname{figure}{Figure}{Figures} %
\usepackage{array}          %
\usepackage{xcolor}
\usepackage{wrapfig}

\definecolor{kellygreen}{rgb}{0.3, 0.73, 0.09}
\definecolor{alizarin}{rgb}{0.82, 0.1, 0.26}

\newcommand{\cmark}{{\color{kellygreen} \ding{51}}}
\newcommand{\xmark}{{\color{alizarin} \ding{55}}}

\usepackage{lipsum}

\usepackage{comment}

\usepackage{csquotes}

\title{Learning to Commit: Generating Organic Pull Requests via Online Repository Memory}

\author{
  \small
  Mo Li$^{1}$ \hspace{0.6em}
  L.H. Xu \hspace{0.6em}
  Qitai Tan$^1$ \hspace{0.6em}
  Ting Cao$^{1\dagger}$ \hspace{0.6em}
  Yunxin Liu$^{1}$ \hspace{0.6em}
  \\
  $^1$Tsinghua University \hspace{1em}
}

\begin{document}

\maketitle
\renewcommand{\thefootnote}{\fnsymbol{footnote}}
\footnotetext[2]{Corresponding author.}

\begin{abstract}
    Large language model (LLM)-based coding agents achieve impressive results on controlled benchmarks yet
    routinely produce pull requests that real maintainers reject.
    The root cause is not functional incorrectness but a lack of \emph{organicity}: generated code ignores
    project-specific conventions, duplicates functionality already provided by internal APIs, and violates
    implicit architectural constraints accumulated over years of development.
    Simply exposing an agent to the latest repository snapshot is not enough: the snapshot reveals the
    final state of the codebase, but not the repository-specific change patterns by which that state was reached.
    We introduce \textbf{Learning to Commit}, a framework that closes this gap through
    \textbf{Online Repository Memory}.
    Given a repository with a strict chronological split, the agent performs
    \emph{supervised contrastive reflection} on earlier commits: it blindly attempts to resolve each historical
    issue, compares its prediction against the oracle diff, and distils the gap into a
    continuously growing set of \emph{skills}---reusable patterns capturing coding style,
    internal API usage, and architectural invariants.
    When a new PR description arrives, the agent conditions its generation on these accumulated skills,
    producing changes grounded in the project's own evolution rather than generic pretraining priors.
    Evaluation is conducted on genuinely future, merged pull requests that could not have been seen
    during the skill-building phase, and spans multiple dimensions including functional correctness,
    code-style consistency, internal API reuse rate, and modified-region plausibility.
    Experiments on an expert-maintained repository with rich commit history show that Online Repository Memory effectively improves organicity scores on held-out future tasks.
    \end{abstract}

\section{Introduction}\label{sec:Introduction}

The past three years have witnessed a remarkable acceleration in AI-assisted software engineering.
Coding agents built on large language models now resolve a substantial fraction of tasks on curated
benchmarks such as SWE-bench \citep{swebench}, while libraries like
HumanEval \citep{humaneval} and MBPP \citep{mbpp} are largely saturated.
Repository-scale benchmarks such as FEA-Bench and FeatureBench further extend evaluation from
isolated code generation to multi-file feature implementation in realistic repositories
\citep{feabench,featurebench}.
These numbers have fuelled optimism that LLM agents are ready for industrial deployment.
Yet professional maintainers of complex repositories remain cautious: they acknowledge that
AI-generated code is often functionally correct, but still routinely reject pull requests because
the code feels \emph{alien}---written by someone who has never read the project. In practice, this
alien quality manifests not only as stylistic mismatch but also as unnecessary code bloat: the agent
re-implements utilities, wrappers, or control flow patterns that the repository already contains.

The gap between benchmark scores and industrial merge-rate reflects a structural blind spot in
how we currently evaluate coding agents. Benchmarks such as SWE-bench still cast software
engineering as a sequence of isolated, one-off tasks: an agent sees an issue, edits the codebase, and
succeeds if the tests pass, without carrying any accumulated knowledge forward.
Furthermore, they typically evaluate issues out of chronological order, ignoring how a repository
and its conventions naturally evolve over time. COMMIT0 \citep{commit0} stresses whole-library implementation from
API specifications, while SWE-CI \citep{sweCI} extends evaluation over months of consecutive
commits. At the same time, repository-level systems such as Repository Memory and SGAgent
show that richer codebase context improves localisation and repair \citep{repoMemory,sgagent},
and PR-centric works such as Coeditor, Clean-PR, and R2E-Gym suggest that commit and PR
histories are valuable learning signals \citep{coeditor,cleanPR,r2egym}. Yet these settings still
share the same basic premise: the agent is not expected to undergo anything like repository-specific
onboarding before it starts writing code.
This omission matters because the current repository snapshot shows only the finished building, not
why particular supports, interfaces, and boundaries were introduced along the way.
More broadly, it reflects a shift that is now becoming central in agent research: capability no
longer depends only on managing context within a single session, but increasingly on managing and
updating memory across sessions and across time.
While recent frameworks like SWE-Bench-CL \citep{swebenchCL} attempt to evaluate such continual
learning chronologically, they rely on unsupervised self-reflection. Without explicit expert
supervision, agents are prone to accumulating self-reinforcing errors.

This missing onboarding step is precisely what separates a seasoned contributor from an outsourced
newcomer. When a human engineer joins a mature codebase, she does not immediately open a text
editor. She reads prior commits, studies module boundaries, discovers which internal utilities are
idiomatic, and internalises the preferences of the maintainers. Only then does she open a pull
request that looks as if it grew organically from the repository itself. Current agents skip this
process entirely, producing what we term \emph{alien code}: syntactically valid, often functionally
correct, but stylistically foreign, architecturally dissonant, and full of redundant reimplementations
of existing helper functions, often inflating patch size in the process. Even when such patches pass
unit tests, maintainers reject them for exactly the reasons that a junior engineer would be asked to
revise a first contribution.

To generate code that is not merely correct but genuinely \emph{organic}, we introduce
\textbf{Learning to Commit}, a framework centred on \textbf{Online Repository Memory}.
The key idea is simple: organicity is learnable from the commit history, which records
how a project has chosen to evolve over time. Given a repository with a strict chronological split,
the agent performs \emph{supervised contrastive reflection} on earlier commits---blindly attempting
each historical change, comparing its prediction against the oracle diff, and distilling the gap
into a continuously growing, incrementally updatable set of \emph{skills}. When a new PR description
arrives, the agent retrieves and conditions its generation on these skills, producing changes aligned
with the repository's naming conventions, preferred abstractions, and maintainer preferences.

Evaluating coding agents is frequently compromised by pre-training data leakage---a
pervasive vulnerability in current evaluations, as evidenced by recent efforts like
SWE-Bench++ \citep{swebenchplusplus}. By design, our same-repository, time-split
evaluation inherently guarantees zero data leakage. Skills
are built exclusively from commits before a hard cutoff date, while evaluation tasks are drawn from
genuinely future merged commits. We measure success not only by functional correctness but
also by code-style consistency, internal API reuse, and modified-region plausibility. While we validate on an expert-maintained internal repository in this work, the pipeline is designed to be extended to high-quality public GitHub repositories in future work. In this paper, we establish repository-personalised adaptation as a
first-class evaluation objective for coding agents. Our contributions are:

\begin{itemize}
    \item We formalise \emph{repository-personalised online adaptation} through the
    \textbf{Learning to Commit} framework, establishing both an \textbf{online learning mechanism} for agents to organically
    align with evolving codebases and a rigorous evaluation paradigm.
    \item We propose a training-free, online skill extraction method that distils repository-specific
    conventions via supervised contrastive reflection---comparing the agent's blind attempts against
    oracle diffs to accumulate abstract, reusable development \emph{skills}.
    \item We construct and release a strict time-split benchmark of curated repositories with
    multi-dimensional metrics (code style, API reuse, modified-region plausibility), demonstrating
    that our framework effectively improves organicity over existing paradigms.
\end{itemize}

\begin{table}[!t]
    \centering
    \small
    \caption{Comparison of representative repository-level benchmarks and agentic learning paradigms. Columns progress from basic capabilities to our unique contributions: leveraging historical context, continual learning across tasks, strict data leakage prevention, oracle-supervised memory updates, and multi-dimensional organicity evaluation. Symbol legend: \cmark\ = fully satisfies; \xmark\ = not supported.}\label{tab:benchmark_positioning}
    \resizebox{\textwidth}{!}{%
    \begin{tabular}{lccccc}
    \toprule
    \textbf{Framework / Paradigm} & \textbf{Chronological Eval} & \textbf{Historical Memory} & \textbf{Zero Data Leakage} & \textbf{Oracle Supervision} & \textbf{Organicity Eval} \\
    \midrule
    SWE-bench \citep{swebench} & \xmark & \xmark & \xmark & \xmark & \xmark \\
    SWE-Bench++ \citep{swebenchplusplus} & \xmark & \xmark & \cmark & \xmark & \xmark \\
    RepoMem \citep{repoMemory} & \xmark & \cmark & \xmark & \xmark & \xmark \\
    SWE-CI \citep{sweCI} & \cmark & \xmark & \xmark & \xmark & \xmark \\
    SWE-Bench-CL \citep{swebenchCL} & \cmark & \cmark & \xmark & \xmark & \xmark \\
    \midrule
    \textbf{Learning to Commit (Ours)} & \cmark & \cmark & \cmark & \cmark & \cmark \\
    \bottomrule
    \end{tabular}%
    }
    \end{table}

\section{Related Work}

\subsection{Static Evaluation Paradigms and Repository-Level Agents}

Early benchmarks evaluated large language models in isolated, stateless environments \citep{humaneval, mbpp}, which recently evolved into realistic, repository-level tasks like SWE-bench \citep{swebench} and its multi-file or generative extensions \citep{feabench, featurebench, commit0}. To tackle these complex scenarios, recent works scale up training data through automated PR harvesting \citep{cleanPR, scaleswe, r2egym}, with works like SWE-Bench++ \citep{swebenchplusplus} specifically designing recent cutoffs to ensure zero data leakage and design multi-agent workflows with specialized roles and reasoning steps \citep{sgagent, agyn}. However, these data-centric and workflow-based approaches treat tasks as static snapshots, completely ignoring the chronological evolution of software. While SWE-CI \citep{sweCI} introduces a temporal CI-loop to penalize technical debt, it primarily evaluates performance degradation rather than empowering the agent to actively learn and internalize repository-specific conventions over time. Crucially, these existing paradigms rely almost exclusively on functional test passes as their success metric, lacking multi-dimensional organicity evaluation to assess codebase stylistic consistency, internal API reuse, and architectural fit.

\subsection{Agent Memory and Continual Learning in Software Engineering}

To capture historical developer intent, models have incorporated commit histories via static weight updates \citep{commitbart, coeditor} or passive retrieval augmentation \citep{repoMemory}, yet both lack an active trial-and-error learning process. The necessity for dynamic continual learning is highlighted by the performance degradation observed in long-horizon tasks \citep{sweCI, sweContextBench}, prompting explorations into online agent memory at various granularities \citep{agentMemorySurvey, subtaskMemory, liveEvo}. Most notably, SWE-Bench-CL \citep{swebenchCL} evaluates continual learning chronologically but relies on unsupervised self-reflection; consequently, it is highly vulnerable to \enquote{garbage-in-garbage-out} errors when early attempts fail, as autonomous reflection without ground-truth validation often leads to self-reinforcing errors \citep{agentMemorySurvey}. Even with stepwise environmental feedback \citep{openclawRL}, signals remain too noisy to extract deep design patterns. Our work bridges this gap: by strictly partitioning history to prevent data leakage, our agent attempts historical commits and receives \emph{oracle diffs} as dense supervision. Through supervised contrastive distillation between blind attempts and expert patches, the agent extracts reusable, repository-specific development patterns into a continuously refined skill document, ensuring failures drive optimal, history-conditioned memory updates.
Table~\ref{tab:benchmark_positioning} summarises how our protocol compares with representative prior benchmarks across these key dimensions.

\section{Methodology}\label{sec:method}

\begin{figure}[!t]
    \centering
    \includegraphics[width=\textwidth]{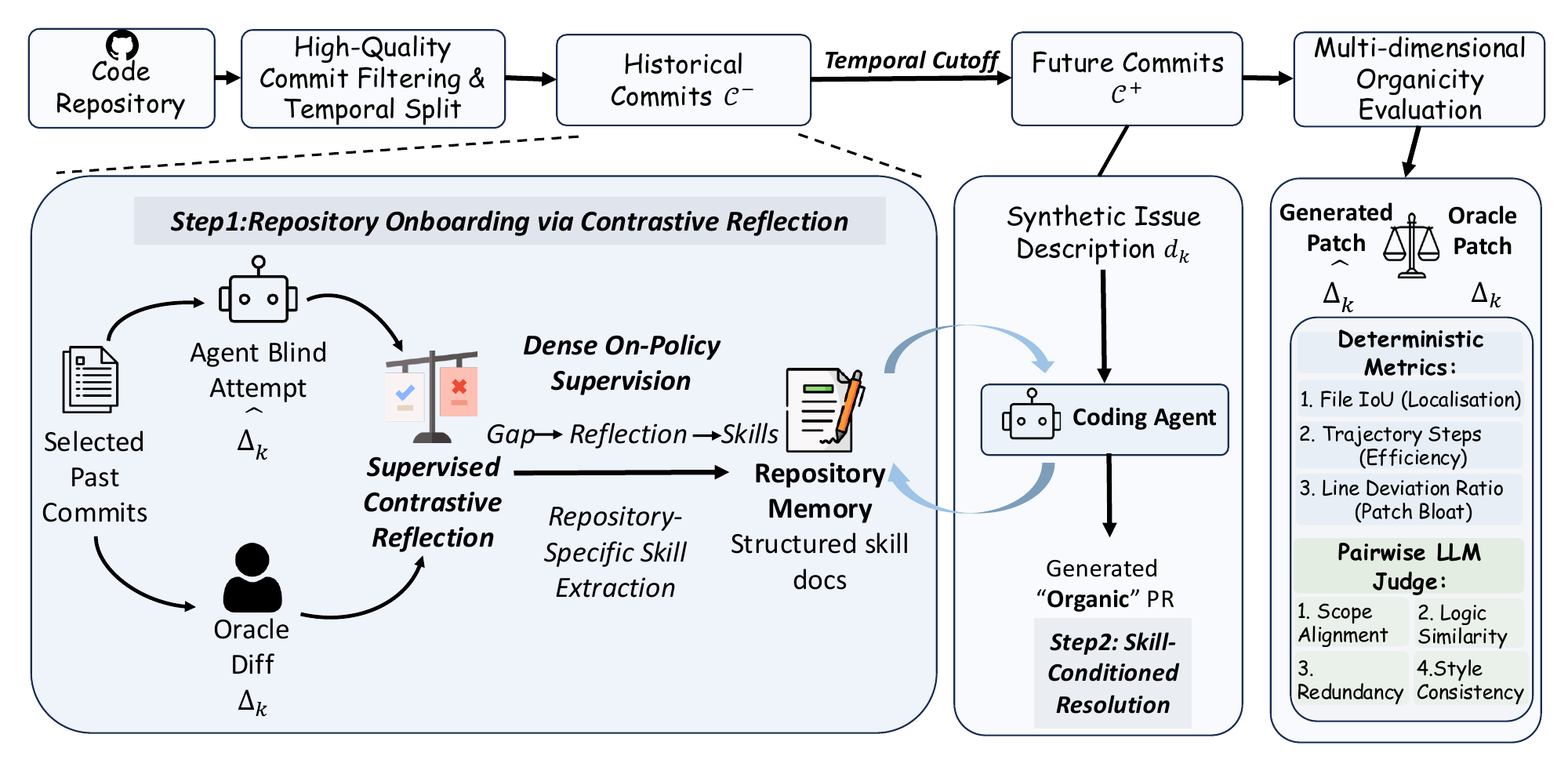}
    \caption{Overview of the Learning to Commit framework. Real repositories are time-split into past commits (for skill learning) and future commits (for evaluation). In Step~1, the agent performs on-policy contrastive reflection: it blindly attempts each historical commit, then compares its prediction against the oracle diff to extract repository-specific development patterns into a structured skill document. In Step~2, the agent generates organic patches for unseen future tasks, guided by the accumulated skills. In Step~3, generated patches are evaluated against oracle commits across multiple dimensions including file IoU, line deviation, and multi-dimensional LLM judge scores.}\label{fig:framework}
\end{figure}

\subsection{Problem Formulation}\label{sec:method:formulation}

Let $\mathcal{C} = (c_1, c_2, \ldots, c_T)$ denote a chronological sequence of high-quality commits from a target repository $\mathcal{R}$. Each commit $c_t$ is associated with a repository snapshot $\mathcal{S}_t$ (the codebase state at the parent commit), an oracle code diff $\Delta_t$, and a synthetic issue description $d_t$ that specifies the task intent without leaking the implementation. A temporal cutoff $T^*$ strictly partitions the sequence into a \emph{history prefix} $\mathcal{C}^- = \{c_t : t \leq T^*\}$ for learning and a \emph{held-out test set} $\mathcal{C}^+ = \{c_k : k > T^*\}$ for evaluation.

Given a future task description $d_k \in \mathcal{C}^+$, the objective is to generate a patch $\hat{\Delta}_k$ that is simultaneously functionally correct and organically aligned with the repository, under the constraint that the agent may only use $\mathcal{C}^-$ for adaptation. The Learning to Commit framework decomposes this into two phases (illustrated in \cref{fig:framework}): (1)~\textbf{Repository Onboarding (Learning Phase)}, where the agent iteratively builds a reusable skill document $\mathcal{M}$ from $\mathcal{C}^-$ through on-policy contrastive reflection; and (2)~\textbf{Skill-Conditioned Resolution (Solve Phase)}, where, given $d_k$ and the repository snapshot $\mathcal{S}_k$, the agent autonomously resolves the task conditioned on the accumulated skills $\mathcal{M}$.

\subsection{Repository Onboarding via Contrastive Reflection}\label{sec:method:onboarding}

The learning phase mimics how a human developer onboards onto a new project: not by passively reading documentation, but by actively attempting tasks and learning from the gap between one's own output and expert practice. We initialise an empty skill document $\mathcal{M}^{(0)} = \varnothing$. For each learning commit $c_t \in \mathcal{C}^-$, the agent executes a three-step loop. \textbf{Step~1 (Blind Attempt):} The agent receives the repository snapshot $\mathcal{S}_t$ and the synthetic issue description $d_t$, along with the current skill document $\mathcal{M}^{(t-1)}$. It autonomously explores the codebase using standard tool-use capabilities (file reading, searching, editing) and produces a candidate patch $\hat{\Delta}_t$. \textbf{Step~2 (Oracle Revelation and Contrastive Reflection):} The ground-truth oracle diff $\Delta_t$---representing the accepted solution by a human domain expert---is revealed. The agent compares its own attempt $\hat{\Delta}_t$ against $\Delta_t$, identifying discrepancies in file localisation, implementation logic, API usage, and coding style. The gap between the two serves as dense, on-policy supervision: the larger the discrepancy, the richer the learning signal. \textbf{Step~3 (Skill Update):} Based on the contrastive reflection, the agent updates the skill document through explicit CRUD operations---creating new entries for previously unknown patterns, revising entries that were partially correct, and deprecating entries contradicted by the oracle evidence:
\begin{equation}
    \mathcal{M}^{(t)} = \textsc{Update}\!\left(\mathcal{M}^{(t-1)},\ \hat{\Delta}_t,\ \Delta_t,\ d_t\right).
    \label{eq:update}
\end{equation}
The resulting skill document $\mathcal{M}$ consolidates abstract, reusable development patterns that typically encompass: (1)~coding style and naming conventions, (2)~the existence and correct usage of internal API utilities, (3)~implicit architectural constraints and module boundaries, and (4)~maintainer preference patterns such as error handling style and test organisation. Unlike static RAG over commit histories, this on-policy learning loop ensures the extracted patterns are precisely calibrated to the agent's own capability gaps---addressing exactly the mistakes the agent would otherwise make.

\subsection{Skill-Conditioned Resolution}\label{sec:method:solve}

When a future task $d_k \in \mathcal{C}^+$ arrives, the agent receives the repository snapshot $\mathcal{S}_k$, the task description $d_k$, and the full accumulated skill document $\mathcal{M}$. The agent then autonomously resolves the task using standard tool-use capabilities---reading relevant files, searching the codebase, and applying edits---conditioned on the development patterns recorded in $\mathcal{M}$. No rigid retrieval pipeline or pre-planned workflow is imposed; the agent decides which skills to consult and which files to explore based on its own judgement, mirroring how a human developer with internalised project knowledge would approach a new task.

\subsection{Dataset Construction}\label{sec:method:benchmark}

We preliminarily validate the framework on a large-scale internal reinforcement learning training repository, where every commit has been reviewed and approved by domain experts, ensuring high-quality ground-truth patches. The data curation pipeline proceeds in five stages. (i)~\textbf{Commit scanning and quality filtering}: we extract the full non-merge commit history, apply programmatic pre-filters to remove trivial changes (e.g., fewer than 10 modified lines, version bumps), and use an LLM to assess whether each remaining commit exhibits substantive, learnable development patterns. Commits whose diffs exceed 180K tokens are excluded. In our pilot repository, this substantially reduces the raw commit pool to high-quality candidates (concrete numbers in \S\ref{sec:experiments}). (ii)~\textbf{Unsupervised category clustering}: to prevent the learning curriculum from collapsing onto a single change pattern, we sample the rationales from the first stage and perform unsupervised clustering to identify core development categories (yielding seven categories spanning architecture design, concurrency, testing, etc.). (iii)~\textbf{Category tagging}: each retained commit is then assigned a primary category label by an LLM that considers the commit title, the rationale, and the full patch content.

Following the initial tagging, we construct the final benchmark sets. (iv)~\textbf{Stratified sampling with temporal split}: commits are chronologically ordered and split into learning and test pools. Within each pool, we perform stratified proportional downsampling across categories to construct a balanced curriculum. (v)~\textbf{Synthetic query generation}: for each task, an LLM synthesises an issue-style natural-language query from the commit message and diff. The prompt specifies the \enquote{what} and \enquote{why} of the task while strictly omitting implementation details (exact file paths, function names, or solution strategies), emulating a realistic user issue that does not leak the oracle patch. Crucially, the strict temporal split ensures that all learning commits predate all test commits, completely preventing information leakage from evaluation tasks into the skill-building phase. Scaling this pipeline to diverse open-source GitHub repositories is a direct extension planned for future work.

\subsection{Evaluation Metrics}\label{sec:method:metrics}

Evaluating the organicity of AI-generated patches requires moving beyond functional test passes. We evaluate generated patches $\hat{\Delta}_k$ against oracle patches $\Delta_k$ across two complementary metric families.

\paragraph{Deterministic code metrics.}
(1)~\emph{File IoU}: the Jaccard similarity $|\mathcal{F}(\hat{\Delta}) \cap \mathcal{F}^*_k| / |\mathcal{F}(\hat{\Delta}) \cup \mathcal{F}^*_k|$ between the files modified by the agent and those in the oracle, measuring localisation accuracy.
(2)~\emph{Trajectory steps}: the total number of tool calls during the solve phase, reflecting problem-solving efficiency.
(3)~\emph{Line deviation ratio}: $(|\hat{\Delta}| - |\Delta_k|) / |\Delta_k|$, measuring patch bloat; agents that fail to reuse internal APIs typically produce inflated diffs.

\paragraph{Multi-dimensional LLM judge.}
We employ a pairwise A/B evaluation protocol in which an advanced LLM compares the baseline agent (without skills) against the skill-conditioned agent across four dimensions: (Q1)~\emph{scope alignment}---accuracy of the modified file and function locations; (Q2)~\emph{logic similarity}---proximity to the oracle's core implementation logic; (Q3)~\emph{redundancy and hallucination}---code conciseness and absence of over-engineering, a key indicator of successful skill utilisation; and (Q4)~\emph{code style}---adherence to the repository's native conventions. To reduce single-judge bias, we run evaluations with two independent judge models and report agreement rates.

\section{Experiments}\label{sec:experiments}

\subsection{Experimental Setup}\label{sec:exp:setup}

\paragraph{Repository and dataset curation.}
We evaluate our framework on an internal, expert-maintained reinforcement learning training repository comprising agent environments, judge evaluation, and orchestration subsystems. Starting with 2,738 non-merge commits, we apply the filtering and LLM-assessment pipeline described in \S\ref{sec:method:benchmark}. This yields 386 high-quality, substantive commits (a 77.2\% suitability rate) distributed across seven core development categories, such as architecture design, concurrency and IPC, and defensive programming. Following a strict temporal split, we perform stratified sampling to construct a balanced curriculum of 24 historical learning commits and 7 genuinely future, held-out test tasks.

\paragraph{Experimental design and baselines.}
To isolate the effect of repository-specific onboarding, we evaluate a \emph{skill-conditioned agent} (which receives the accumulated memory $\mathcal{M}$) against a \emph{baseline agent} (which operates without any skill document), both powered by Claude Opus 4.6 with identical tool-use capabilities. We investigate the skill-building process across four experimental conditions by crossing two learning modes with two curriculum assignments. For the learning mode, the agent extracts skills either \emph{Sequentially} (progressively fusing insights through iterative trial-and-error) or in \emph{Parallel} (processing commits concurrently before a final pairwise merge). For curriculum assignment, each test task is paired either with learning commits strictly from its own category (\emph{by\_category}) to simulate targeted onboarding, or with the entire learning corpus (\emph{all}) to simulate comprehensive repository adaptation. This yields four conditions---\texttt{seq-all}, \texttt{seq-bycat}, \texttt{par-all}, and \texttt{par-bycat}---each evaluated on all 7 test tasks.

\subsection{Main Results}\label{sec:exp:results}

\paragraph{Deterministic code metrics.} Table~\ref{tab:hard_metrics} reports file-level localisation accuracy (File IoU), problem-solving efficiency (trajectory steps), and patch bloat (line deviation ratio) across all four conditions. The skill-conditioned agent achieves consistently higher File IoU in three out of four settings, with the largest gain of +19 percentage points in \texttt{seq-all} (80\% vs.\ 61\%). In the same setting, the skill agent also uses 21\% fewer tool calls (56.8 vs.\ 71.9 steps), suggesting that accumulated skills help the agent navigate the codebase more efficiently. Line deviation ratio is lower for the skill agent in three out of four settings, indicating patches closer in size to the oracle.

\begin{table}[t]
\centering
\small
\caption{Deterministic code metrics across four experimental conditions. Bold indicates the better result in each pair. File IoU measures localisation accuracy ($\uparrow$), Steps measures solve efficiency ($\downarrow$), and Line Deviation measures patch bloat relative to the oracle ($\downarrow$, closer to 0 is better).}\label{tab:hard_metrics}
\begin{tabular}{lccc}
\toprule
\textbf{Setting} & \textbf{File IoU} (Skill / Base) & \textbf{Steps} (Skill / Base) & \textbf{Line Dev.} (Skill / Base) \\
\midrule
\texttt{seq-all}    & \textbf{80\%} / 61\% & \textbf{56.8} / 71.9 & \textbf{0.69} / 1.59 \\
\texttt{par-bycat}  & \textbf{80\%} / 68\% & 62.7 / 62.7           & \textbf{0.62} / 0.88 \\
\texttt{par-all}    & \textbf{81\%} / 72\% & \textbf{67.7} / 68.3 & \textbf{0.80} / 1.06 \\
\texttt{seq-bycat}  & 71\% / 71\%           & \textbf{75.0} / 76.4 & 1.22 / \textbf{1.13} \\
\bottomrule
\end{tabular}
\end{table}

\paragraph{Multi-dimensional LLM judge.} Table~\ref{tab:pairwise} reports overall pairwise win rates from two independent judge models (Claude Opus 4.6 and Gemini 3.1 Pro). The skill-conditioned agent wins more often than the baseline in three out of four settings under both judges, with the strongest results in \texttt{par-bycat} (54\%/57\%) and \texttt{seq-all} (55\%/58\%). Table~\ref{tab:dimension_breakdown} breaks down the \texttt{par-bycat} results by dimension: the skill agent's advantage is concentrated in Q2 (logic similarity, 50\% vs.\ 25\%) and Q3 (redundancy reduction, 54\% vs.\ 41\%), while the baseline retains a slight edge on Q1 (scope alignment) and Q4 (code style). Across all four settings, Q3 (redundancy and hallucination) shows the most consistent skill advantage (39--69\% win rate), with high inter-judge agreement.

\begin{table}[t]
\centering
\small
\caption{Overall pairwise win rates (\%) of the skill-conditioned agent, as judged by two independent LLMs. Bold indicates skill agent wins more than 50\%.}\label{tab:pairwise}
\begin{tabular}{lcc}
\toprule
\textbf{Setting} & \textbf{Claude Judge} & \textbf{Gemini Judge} \\
\midrule
\texttt{par-bycat}  & \textbf{54\%} & \textbf{57\%} \\
\texttt{seq-all}    & \textbf{55\%} & \textbf{58\%} \\
\texttt{par-all}    & 46\%           & \textbf{54\%} \\
\texttt{seq-bycat}  & 26\%           & 46\%           \\
\bottomrule
\end{tabular}
\end{table}

\begin{table}[t]
\centering
\small
\caption{Dimension-level breakdown for the \texttt{par-bycat} setting (averaged over Claude and Gemini judges).}\label{tab:dimension_breakdown}
\begin{tabular}{lccc}
\toprule
\textbf{Dimension} & \textbf{Skill Win} & \textbf{Base Win} & \textbf{Tie} \\
\midrule
Q1: Scope Alignment         & 25\% & 41\% & 34\% \\
Q2: Logic Similarity        & \textbf{50\%} & 25\% & 25\% \\
Q3: Redundancy \& Halluc.   & \textbf{54\%} & 41\% & 5\%  \\
Q4: Code Style              & 27\% & 38\% & 36\% \\
\bottomrule
\end{tabular}
\end{table}

\subsection{Analysis}\label{sec:exp:analysis}

\paragraph{Learning mode comparison.} Sequential learning with full-corpus assignment (\texttt{seq-all}) produces the highest-quality skills: File IoU reaches 80\%, trajectory steps are minimised, and pairwise win rate is among the highest. This is consistent with our expectation that progressive, on-policy refinement of a single skill document yields more coherent and deeply fused knowledge than parallel extraction followed by post-hoc merging. However, \texttt{par-bycat} achieves comparable File IoU (80\%) with targeted category-specific skills, suggesting that focused, domain-aligned learning can compensate for the lack of sequential fusion.

\paragraph{The asymmetric value of skills.} Our results reveal that repository-specific skills provide asymmetric benefits across capability dimensions. The strongest gains appear in \emph{file localisation} (File IoU +10--18\%), \emph{core logic reproduction} (Q2 win rate 50\% vs.\ 25\%), and \emph{redundancy reduction} (Q3 favours skill agents in three out of four settings). Meanwhile, fine-grained scope alignment (Q1) and code style (Q4) show marginal or neutral effects. This pattern suggests that accumulated skills primarily help agents understand \emph{where} to make changes and \emph{what} internal patterns to reuse, while surface-level stylistic conformance may require additional mechanisms beyond the current skill representation.

\paragraph{Case study.} The largest positive effect appears on a task requiring the agent to fix a shared-RNG concurrency bug in a judge client module. The skill agent, having learned the repository's module structure during onboarding, correctly localises the fix to the judge client file (File IoU = 100\%), while the baseline agent misidentifies the target entirely, editing an unrelated API module (File IoU = 0\%). Conversely, the largest negative effect occurs on a training-step guard logic fix, where both agents correctly identify the core bug but the skill agent retains a redundant defensive guard that the baseline omits, resulting in slightly more verbose code. This illustrates that skills occasionally induce over-caution without functionally harming the solution.

\section{Conclusion and Future Work}\label{sec:conclusion}

While large language model-based coding agents excel on isolated programming benchmarks, they frequently fail to produce \enquote{organic} pull requests in real-world software engineering because they lack an understanding of repository-specific human preferences and historical context. To bridge this gap, we presented \textbf{Learning to Commit}, an agentic framework that empowers models to actively learn from a repository's evolution. By performing on-policy contrastive reflection on historical commits, the agent distils implicit architectural constraints and coding conventions into a reusable repository memory. Our results demonstrate that this historical adaptation enables agents to significantly improve file localisation, reduce patch bloat, and generate contributions that genuinely align with expert maintainers' expectations.

As a foundational step toward repository-adaptive agents, this work naturally opens several avenues for future research. First, while our framework demonstrates strong efficacy on a highly coupled, industrial codebase, scaling the data curation and evaluation pipeline to a diverse set of high-quality, popular open-source GitHub repositories is a critical next step to verify its generalisability. Second, although our multi-dimensional LLM judge provides a granular assessment of organicity, relying on LLMs as evaluators inherently introduces potential biases and inaccuracies; future work must explore more robust, verifiable metrics for evaluating true alignment with human coding preferences. Finally, an important direction is to examine whether repository-specific skills transfer to broader coding tasks and established evaluation settings, further clarifying the scope and limits of historical adaptation in moving coding agents toward seasoned, long-term contributors.

\bibliographystyle{unsrtnat}
\bibliography{references}

\clearpage
\appendix

\newpage

\end{document}